\documentclass[final,5p,times,twocolumn,authoryear]{elsarticle}

\usepackage{amssymb}
\usepackage{amsmath}
\usepackage{empheq}
\usepackage{comment}
\usepackage{balance}
\usepackage{placeins}
\usepackage{enumerate}
\usepackage[inline,shortlabels]{enumitem}
\usepackage[usenames,dvipsnames]{color}
\usepackage{xcolor}
\usepackage[colorlinks=true,linkcolor=Maroon,citecolor=Maroon,urlcolor=Maroon]{hyperref}
\usepackage[utf8]{inputenc}
\usepackage[english]{babel}
\usepackage{geometry}
\usepackage{booktabs}
\usepackage{array}
\usepackage{longtable}
\usepackage{graphicx}
\usepackage{fancyhdr}
\usepackage{titlesec}
\usepackage{enumitem}
\usepackage{hyperref}
\usepackage{caption}
\usepackage{float}
\usepackage{url}

\begin{document}

\begin{frontmatter}

\title{Case Study of a 75-Year-Old Woman with Parkinson's Disease:\\
Rehabilitation Trajectory with Logic Workout Training}
 
  \author[add1]{Paul-Emmanuel Sornette}
 \author[add1,add2]{Didier Sornette\corref{cor1}}
 
  \address[add1]{\scriptsize
        Logic Workout GmbH
    }
 
 \address[add2]{\scriptsize
        Institute of Risk Analysis, Prediction and Management (Risks-X),
        Academy for Advanced Interdisciplinary Sciences,\\
        Southern University of Science and Technology, Shenzhen, China
    }
    \cortext[cor1]{Corresponding author. Email: dsornette@ethz.ch}
\date{\today}

\begin{abstract}
We report the single-case trajectory of a 75-year-old retired occupational female therapist with idiopathic Parkinson's disease, Hoehn and Yahr stage 2 at diagnosis. Following progressive impairment despite standard care, she initiated training with Logic Workout (LW) in July 2025 under supervision. Within weeks, she reported meaningful improvements spanning motor function, posture, pain, fine motor skill, mood, sleep consolidation, and disappearing of fatigue. Although single cases cannot establish generalizable efficacy, systematic and chronological documentation can be valuable for hypothesis generation and feasibility assessment in real-world settings prior to controlled trials. We summarize the baseline condition and treatment history, describe the LW intervention, compile self-reported outcomes, and interpret the findings in light of the underlying Logic Workout hypothesis, before concluding with key caveats and perspectives for future research.
\vskip 0.3cm
\noindent
{\bf Keywords}: Parkinson's disease; Logic Workout; Dynamic Instability Training; Reactive Falling Effect; Neuromuscular Control; Rehabilitation; Motor Plasticity; Performance Enhancement; Proprioceptive Stimulation; Case Study.
\end{abstract}

\end{frontmatter}

\vspace{1em}

\section{Introduction and Rationale}

This report presents the trajectory of a patient with Parkinson's disease who exhibited a striking recovery after initiating the Logic Workout (LW) instability-based training method. 

Such single-case documentation is valuable because it captures the complexity of individual responses, including motivation, proprioception, and non-motor symptoms that are often missed in standardized group studies. In neurorehabilitation and movement disorders, longitudinal, naturalistic observation remains essential for understanding functional and qualitative aspects of recovery. Although randomized controlled trials are the gold standard for establishing efficacy, detailed single-case studies continue to play a crucial role in medical discovery: they illuminate mechanisms, refine hypotheses, and guide the design of future controlled investigations \citep{Yin2017,Flyvbjerg2006,Green2018}.

\section{History and Baseline Condition}
Symptoms consistent with Parkinsonism first appeared in 2023 progressively developing into a persistent left-sided resting tremor, bradykinesia\footnote{\textbf{Bradykinesia} refers to the slowness of voluntary movement characteristic of Parkinson's disease. It involves delayed initiation, reduced amplitude, and progressive slowing of repetitive or sequential motor actions.}, and diminished arm swing by the end of that year. 
In March 2024, the Neurology Department of Kantonsspital St.\ Gallen confirmed idiopathic \footnote{\textbf{Idiopathic} means that the condition arises spontaneously or from an unknown cause, without evidence of a secondary or hereditary origin. In medicine, it denotes diseases with no identifiable underlying factor.} Parkinson's disease, Hoehn and Yahr stage 2 \footnote{\textbf{Hoehn and Yahr stage 2} refers to the second level of the Hoehn and Yahr scale, which classifies the severity of Parkinson's disease based on motor symptoms and functional impairment. Stage 2 indicates bilateral or midline involvement (affecting both sides of the body) without impairment of balance. Patients at this stage typically show mild bradykinesia, rigidity, or tremor on both sides but remain physically independent and able to maintain upright posture.}, with a strongly positive family history, as her father, grandfather, and aunt had also been affected \citep{KSSG2024}.  
Levodopa/benserazide (Madopar) \footnote{\textbf{Levodopa/benserazide (Madopar)} is a combination medication used to treat Parkinson's disease. Levodopa serves as a precursor to dopamine, replenishing the brain's deficient neurotransmitter, while benserazide inhibits peripheral decarboxylase enzymes, preventing premature conversion of levodopa to dopamine outside the brain and thereby reducing side effects such as nausea.} was introduced with a gradual titration plan and provided partial symptomatic relief.  
The patient also suffered from distal sensory-predominant polyneuropathy \footnote{\textbf{Distal sensory-predominant polyneuropathy} is a disorder of the peripheral nerves that primarily affects sensory fibers in the distal (farther from the center) parts of the limbs, such as the feet and hands. It typically causes numbness, tingling, or burning sensations and may impair balance or coordination. The term ``predominant'' indicates that sensory symptoms are more pronounced than motor weakness.}
probably related to type 2 diabetes and prior R-CHOP chemotherapy \footnote{\textbf{R-CHOP chemotherapy} is a standard combination regimen used to treat non-Hodgkin lymphomas. The acronym stands for Rituximab, Cyclophosphamide, Hydroxydaunorubicin (also known as Doxorubicin), Oncovin (Vincristine), and Prednisone. Rituximab is a monoclonal antibody targeting CD20 on B lymphocytes, while the other agents are cytotoxic drugs that act synergistically to destroy malignant cells.} for diffuse large B-cell lymphoma \footnote{\textbf{B-cell lymphoma} is a type of cancer that arises from B lymphocytes, the white blood cells responsible for producing antibodies as part of the immune system. In this malignancy, abnormal B cells grow uncontrollably, often forming tumors in lymph nodes or other organs. Diffuse large B-cell lymphoma is the most common aggressive subtype of non-Hodgkin lymphoma.} in 2018, adding sensory imbalance to her motor limitations.

A second outpatient evaluation in April 2024 at Rehaklinik Zihlschlacht confirmed the diagnosis and documented hypomimia \footnote{\textbf{Hypomimia} refers to a reduction or loss of facial expressiveness, often described as a ``masked face.'' It results from decreased spontaneous and voluntary facial muscle movements and is a common motor feature of Parkinson's disease.}, mild rigidity of the left arm, bradykinesia, unsteady broad-based gait, and mild postural instability, while cognition and affect were preserved \citep{Rehaklinik2024}.  
An inpatient rehabilitation course was advised, but persistent fatigue, back pain, and limited mobility continued through 2025 despite physiotherapy. 

Before initiating Logic Workout training, the patient had been suffering from chronic lower-back pain, a long-standing condition severe enough to require medical intervention. Two months prior to beginning the LW program, she had received two infiltrations, in the form of targeted injections administered into the spinal joints to reduce inflammation and alleviate pain. Despite these infiltrations, her pain persisted, indicating that the underlying problem had not been resolved.

The patient had also been undergoing weekly physiotherapy, which provided short-term relief lasting approximately two days. This pattern of transient improvement followed by symptom recurrence was documented over multiple treatment cycles. Recognizing the limited durability of this approach, she discontinued physiotherapy and initiated Logic Workout training. This report documents the factual outcomes of that transition.

The patient's medical background includes type 2 diabetes mellitus \footnote{\textbf{Type 2 diabetes mellitus} is a chronic metabolic disorder characterized by elevated blood glucose levels due to insulin resistance and an inadequate compensatory insulin response. Over time, it can lead to complications affecting the cardiovascular system, kidneys, eyes, and peripheral nerves.}
 (since 2002), hypertensive heart disease, obesity, obstructive sleep apnea, bilateral knee replacements (2022), breast carcinoma (2021) treated surgically with ongoing endocrine therapy, and the above-mentioned lymphoma in remission since 2018.  
By mid-2025, she reported being largely unable to walk unassisted and described a sensation of being ``glued to the ground,'' with whole-body tremor and loss of fine motor control.

\section{Intervention: Logic Workout Training}

Beyond its mechanical simplicity, the Logic Workout system is grounded in a neurobiological model of \emph{reactive instability training}, known as the ``Reactive Falling Effect'' \citep{SSLW2025}. This paradigm proposes that controlled exposure to small, multidirectional instabilities, such as those induced by the elastic deformations and rolling motions of the fitball, activates rapid compensatory responses within the spinal and supraspinal circuits responsible for balance and postural control. These perturbations induce subthreshold episodes of ``reactive falling,'' brief instances where the center of mass drifts toward instability and triggers near-reflexive corrections mediated by proprioceptive and vestibular feedback loops.

The patient first learned about Logic Workout through the authors' scientific manuscript ``Harnessing the Reactive Falling Effect for Rehabilitation and Performance Boosting'' \citep{SSLW2025}, which had been publicly shared on the international research repository \url{arxiv.org}. After reading the sections on brain activation and neurophysiological mechanisms, she recognized their potential relevance to her own condition and contacted the authors to request specific guidance adapted to her severely limited mobility.

The first author conducted multiple video consultations to assess her baseline abilities and design a customized progression plan. While the original LW protocol was developed for healthy users and athletes, the training was carefully adapted and extended to accommodate the patient's specific condition, which included marked motor handicap, obesity, postural instability, and a history of multiple comorbidities. All exercises were adapted to be performed in a seated position, ensuring safety while still providing the proprioceptive challenge central to the LW methodology. 
The patient adjusted the frequency according to her needs, eventually performing two 10-minute sessions per day, about five days per week. 
As her condition improved rapidly over subsequent weeks, additional exercises were progressively introduced. 

By late August, she had progressed from seated exercises to standing positions, incorporating dynamic transitions and balance challenges, ultimately including single-leg stance tasks on the ball for up to 45\,s, with exercises documented via the Logic Workout mobile application.

\section{Outcomes and Functional Milestones}

The patient's progress notes from July through September 2025 document a sequence of functional milestones.

\subsection*{Early Phase (Weeks 1--2)}
Within two weeks of starting LW, the patient reported inner calm, improved mood, greater trunk mobility, and complete disappearance of chronic lower-back pain. The pain resolution occurred without any additional medical treatment.

\subsection*{Weeks 3--6}
By mid-August 2025, she could stand and cook for 90 minutes, resumed evening social activities, and described a return of fine manual dexterity, with her ``hands and fingers moving normally again.'' She successfully addressed 180 envelopes, a task previously impossible due to finger stiffness. Morning routines, including dressing, became smoother, and she reported feeling relaxed throughout the day.

She also noted a marked reduction in postural sway and increased confidence in weight-shifting maneuvers. The sensation of being ``glued to the ground,'' characteristic of freezing-of-gait phenomena in Parkinson's disease, diminished substantially. She regained the ability to perform fine motor tasks such as peeling potatoes and cutting vegetables, activities that had become impossible due to tremor and coordination deficits.

\subsection*{Weeks 7--8}
On 26 August 2025, she reported walking freely without assistance on uneven ground, sleeping six to seven hours continuously, and experiencing clearer daytime concentration. Sleep architecture improved, with fewer nocturnal awakenings and more consolidated rest periods. By mid-August, she could perform exercises standing on the fitball, sustaining postural control for extended durations. Tremor amplitude appeared subjectively reduced, and handwriting became more legible.

\subsection*{Weeks 9--12}
By early September, she could climb narrow staircases unaided and stay socially active until midnight without mental fog. Her use of assistive devices decreased, and she moved freely within familiar environments. On 10 September 2025, she attended an endocrinology appointment alone for the first time in seven years, noting complete absence of tremor during the visit. Although lunge exercises remained demanding, one-foot balance on the ball improved toward 45\,s, and morning support requirements declined.

\subsection*{Third-Party Medical Professional Observations}

During this period, the patient attended a follow-up appointment with her orthopedic surgeon, the physician who had performed both of her knee replacement surgeries and was therefore familiar with her baseline condition and medical history. The orthopedist observed her gait and movement patterns. According to the patient's report of the consultation, the physician stated that an observer would not notice that she has Parkinson's disease. This assessment occurred under non-optimal conditions: the elevator was out of service, requiring her to climb one flight of stairs while under the stress of the appointment setting. The orthopedist documented steady walking despite these circumstances.

In late August 2025, the patient's regular physiotherapist, who had been providing treatment before the start of Logic Workout training, observed her marked functional improvements and suggested reducing the session frequency to once every two weeks. The patient herself questioned the continued necessity of conventional physiotherapy, noting that despite years of regular sessions, previous exercises had produced only short-lived benefits, typically lasting no more than two days, whereas the effects experienced with Logic Workout appeared more sustained and comprehensive.

\section{Contextual Interpretation of the Results}

The patient's case illustrates a rapid and broad functional recovery following initiation of the Logic Workout (LW) program. Within eight weeks, she demonstrated sustained improvements in posture, balance, fine dexterity, mobility, and chronic pain. These outcomes stand in marked contrast to the limited, short-lived benefits previously achieved after years of conventional physiotherapy.
These changes were independently observed by healthcare professionals familiar with her condition and temporally coincided with the introduction of LW training, suggesting a meaningful causal contribution.

Parkinson's disease is primarily characterized by motor symptoms resulting from dopaminergic depletion\footnote{\textbf{Dopaminergic depletion} refers to the progressive loss of dopamine-producing neurons, mainly in the substantia nigra pars compacta of the midbrain. Dopamine is essential for smooth and coordinated movement; its deficiency disrupts basal-ganglia signaling, producing tremor, rigidity, and bradykinesia.} in the basal ganglia, but its effects extend to cortical, cerebellar, and proprioceptive circuits.
Standard pharmacotherapy alleviates neurotransmitter imbalance but does not retrain sensorimotor coordination, making physical therapy a cornerstone of non-pharmacological management \citep{Ernst2023}.

Recent meta-analyses confirm that task-specific and balance-oriented training yields superior outcomes compared with passive or low-intensity interventions \citep{Wang2023}.
In particular, perturbation-based and reactive-balance exercises engage anticipatory and compensatory mechanisms that are crucial for fall prevention and gait stability.

Instability-based approaches, such as Logic Workout, are consistent with this body of evidence: they deliver a safe yet challenging sensorimotor stimulus that recruits postural reflexes and promotes adaptive control.
Mechanistically, dynamic-instability practice may strengthen anticipatory postural adjustments, enhance inter-limb coordination, and re-engage dormant motor circuits, paralleling findings that intensive balance training drives neuroplastic reorganization in Parkinson's disease \citep{Ernst2023,Wang2023,Lee2016}.

\section{Mechanistic Interpretation within the Theoretical Framework}

The observed recovery aligns closely with the theoretical model developed by \citet{SSLW2025}, which introduces the concept of the Reactive Falling Effect as the central mechanism of Logic Workout.

According to this framework, repeated exposure to controlled micro-instabilities, brief sub-threshold ``near-falls'', elicits rapid corrective responses that simultaneously activate spinal reflex arcs, cerebellar feedback loops, and cortical planning networks.
Each perturbation functions as a small but asymmetric input that transiently amplifies neural activity within the motor-control hierarchy, producing what the authors term reactive neural instability-driven amplification.

This transient amplification acts as a form of structured physiological ``noise,'' enhancing responsiveness and tuning the gain of sensorimotor feedback without destabilizing the system.
Through numerous, progressively varied repetitions of micro-instabilities, these micro-instabilities promote directed neuroplasticity, reinforcing the coupling between predictive (feedforward) and corrective (feedback) control mechanisms.
In effect, the nervous system learns to anticipate and counteract perturbations more efficiently, leading to improved balance, agility, and postural confidence.

In the patient's case, the recovery of balance, dexterity, and walking autonomy suggests successful 
reactivation of both fast cerebellar-spinal reflex control and slower adaptive recalibration mediated by cortical and basal-ganglia pathways.
Given that Parkinson's disease involves not only dopaminergic loss but also a breakdown in the dynamic coordination among these subsystems, the Logic Workout intervention appears to have partially re-synchronized them through repeated proprioceptive perturbation and real-time correction.

While this remains a single observation, it exemplifies how the Reactive Falling Effect can transform instability, a symptom typically feared in Parkinson's disease, into a therapeutic resource.
By turning controlled imbalance into structured training, the method re-engages sensorimotor integration and demonstrates, in this real-world case, the potential validity of the theoretical framework proposed by \citet{SSLW2025}.
These results encourage formal clinical trials to determine whether the same mechanisms can be harnessed systematically to enhance rehabilitation outcomes across neurological and musculoskeletal disorders.

\section{Future Directions and Ethical Note}

The patient's professional background as an occupational therapist may have supported adherence and focused engagement, potentially amplifying the intervention's effects.
Although this is a single case without quantitative gait analysis or control comparison, the durability and breadth of improvement, together with prior evidence supporting reactive-balance paradigms, justify systematic follow-up studies to confirm and quantify the therapeutic potential of the LW program.

Future investigations should employ prospective cohort or crossover designs comparing LW with standard physiotherapy, using validated clinical scales (UPDRS \footnote{\textbf{UPDRS (Unified Parkinson's Disease Rating Scale)} is a standardized clinical scale used to quantify the severity and progression of Parkinson's disease. It evaluates motor and non-motor symptoms across several domains, including mentation, activities of daily living, motor examination, and treatment-related complications. The motor subscale (Part III) is most often used to assess therapeutic efficacy.}, Mini-BESTest \footnote{\textbf{Mini-BESTest (Mini Balance Evaluation Systems Test)} is a shortened version of the Balance Evaluation Systems Test designed to assess dynamic balance control. It measures anticipatory and reactive postural adjustments, sensory orientation, and gait stability, providing a sensitive tool for evaluating balance deficits and rehabilitation outcomes in neurological conditions such as Parkinson's disease.}), quantitative postural sway analysis, and wearable motion sensors to capture objective biomarkers of improvement.  
Integrating digital adherence data within the LW application would support scalability and safety monitoring.  

The patient provided written informed consent for publication of her clinical course and associated data, with all identifying details minimized in accordance with ethical standards.

\section*{Acknowledgments}
We sincerely thank the patient for her trust, perseverance, and permission to share her experience. We are also deeply grateful to her and her husband for reviewing this report and confirming the factual accuracy of its content.

\section*{Conflict of Interest}
Logic Sports GmbH is the developer of the Logic Workout technology used in this report.

\bibliographystyle{elsarticle-harv}

\end{document}